\PassOptionsToPackage{table,dvipsnames}{xcolor}
\documentclass[acmtog,nonacm]{acmart}

\usepackage{booktabs} 
\usepackage{bm}

\usepackage[ruled]{algorithm2e} 

\SetAlFnt{\small}
\SetAlCapFnt{\small}
\SetAlCapNameFnt{\small}
\SetAlCapHSkip{0pt}
\usepackage{multirow}
\usepackage{makecell}
\usepackage[makeroom]{cancel}

\usepackage{calc}

\definecolor{tabfirst}{rgb}{1, 0.7, 0.7}
\definecolor{tabsecond}{rgb}{1, 0.85, 0.7}
\definecolor{tabthird}{rgb}{1, 1, 0.7}
\definecolor{tabgray}{rgb}{0.9, 0.9, 0.9}
\definecolor{turquoise}{cmyk}{0.65,0,0.1,0.3}
\definecolor{purple}{rgb}{0.65,0,0.65}
\definecolor{darkgreen}{rgb}{0, 0.5, 0}
\definecolor{orange}{rgb}{0.8, 0.6, 0.2}
\definecolor{red}{rgb}{0.8, 0.2, 0.2}
\definecolor{darkred}{rgb}{0.6, 0.1, 0.05}
\definecolor{blueish}{rgb}{0.0, 0.3, .6}
\definecolor{light_gray}{rgb}{0.7, 0.7, .7}
\definecolor{pink}{rgb}{1, 0, 1}
\definecolor{greyblue}{rgb}{0.25, 0.25, 1}

\newcommand{\figref}[1]{Fig.~\ref{fig:#1}}
\newcommand{\tabref}[1]{Table~\ref{tab:#1}}

\newcommand{\secref}[1]{Sec.~\ref{sec:#1}}

\newcommand\lft{\mathopen{}\left}
\newcommand\rgt{\aftergroup\mathclose\aftergroup{\aftergroup}\right}

\newcommand{\ablationMLP}{w/o MLP}
\newcommand{\ablationProbes}{w/o multiple probes}
\newcommand{\ablationExposure}{w/o decreasing exposure}

\newcommand{\image}{I}
\newcommand{\maskBall}{M^B}
\newcommand{\depthBall}{D^{I,B}}
\newcommand{\exposure}{\textit{ev}}
\newcommand{\denoiseNet}{\epsilon_\theta}
\newcommand{\emb}{\zeta}
\newcommand{\mlp}{L_\psi}

\acmYear{2025}
\begin{document}

\title{Spatiotemporally Consistent Indoor Lighting Estimation with Diffusion Priors}
\author{Mutian Tong}
\affiliation{
  \institution{Columbia University}
  \city{New York}
  \country{USA}
}
\email{mt3566@columbia.edu}
\author{Rundi Wu}
\affiliation{
  \institution{Columbia University}
  \city{New York}
  \country{USA}
}
\author{Changxi Zheng}
\affiliation{%
  \institution{Columbia University}
  \city{New York}
  \country{USA}
}
\makeatletter
\let\@authorsaddresses\@empty
\makeatother

\begin{teaserfigure}
    \centering
    \includegraphics[width=0.99\linewidth]{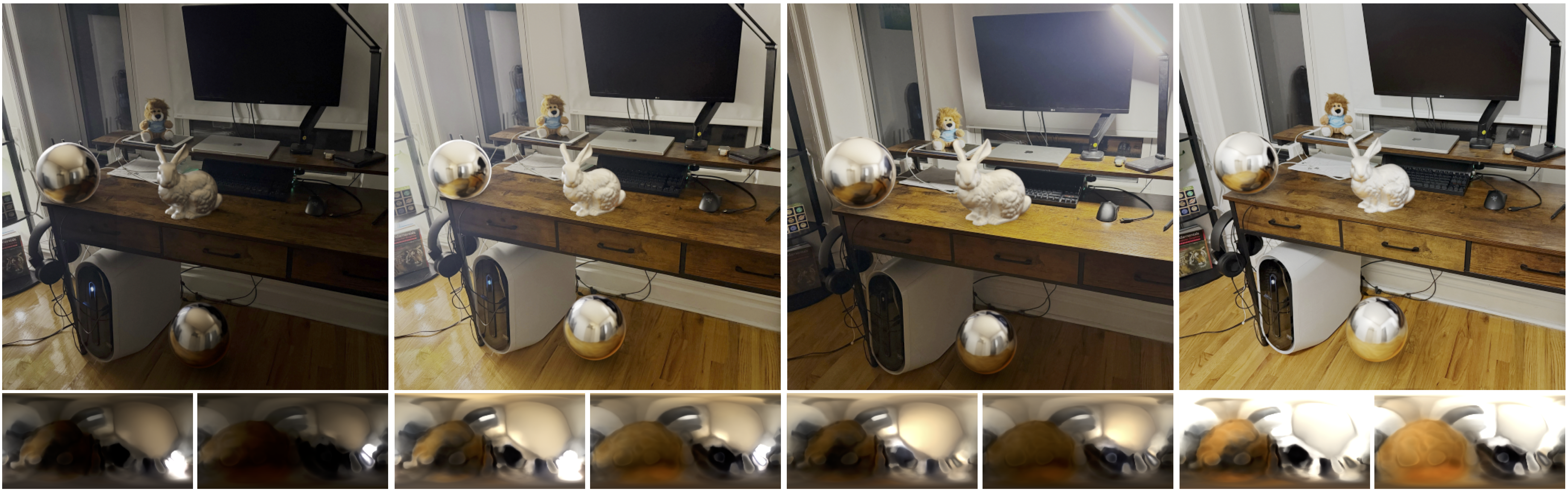}
    \vspace{-2pt}
    \caption{We present a method that achieves consistent indoor lighting estimation from a video wherein the lighting condition changes spatially and temporally. Here we show four frames of an video. For each frame, we show \textbf{(top)} the result of virtual object insertion under the estimated lighting and \textbf{(bottom)} estimated environment maps at locations where the two mirror spheres are inserted. In each column, the environment map on the bottom-left corresponds to the mirror sphere above the desk, and the one on the bottom-right corresponds to the sphere under the desk.
    }
    \label{fig:teaser}
\end{teaserfigure}

\begin{abstract}
Indoor lighting estimation from a single image or video remains a challenge due to its highly ill-posed nature, especially when the lighting condition of the scene varies spatially and temporally.
We propose a method that estimates from an input video a continuous light field describing the spatiotemporally varying lighting of the scene.
We leverage 2D diffusion priors for optimizing such light field represented as a MLP.
To enable zero-shot generalization to in-the-wild scenes, we fine-tune a pre-trained image diffusion model to predict lighting at multiple locations by jointly inpainting multiple chrome balls as light probes.
We evaluate our method on indoor lighting estimation from a single image or video and show superior performance over compared baselines.
Most importantly, we highlight results on spatiotemporally consistent lighting estimation from in-the-wild videos, which is rarely demonstrated in previous works.
\end{abstract}

%
%

%
%


\maketitle

\section{Introduction}
\label{sec:introduction}

High-quality lighting is crucial for virtual object insertion in applications such as augmented reality and video composition. Oftentimes, in these applications, one needs to estimate lighting represented as an environment map from a single image or video of the scene. However, this problem remains challenging due to its highly ill-posed nature\textemdash the estimated environment map must have a high dynamic range (HDR) and capture the scene content beyond the narrow field of view of the input low dynamic range (LDR) images. The problem becomes even more challenging when the input video captures an indoor scene where the lighting condition changes spatially and temporally: lighting intensity may vary as one moves around, and some lights may be turned on or off over time.

Several existing works attempt to estimate HDR lighting from LDR image(s) of a \emph{static} scene. Some can only predict a global illumination of the scene, ignoring spatially varying lighting effects~\cite{barron2014shape,yu2019inverserendernet,legendre2019deeplight,wang2022stylelight,dastjerdi2023everlight,Phongthawee2023DiffusionLight}, while others estimate spatially varying lighting by predicting either an environment map per pixel~\cite{garon2019fast,li2020inverse,li2021openrooms,zhu2022irisformer} or a 3D lighting volume~\cite{srinivasan2020lighthouse,wang2021learning,li2023spatiotemporally,wang2024lightoctree}.
Yet, none of the existing methods can estimate \emph{spatio-temporally consistent} lighting from a video that captures a dynamic lighting condition.

In this work, we present a method that estimates from an input video a continuous, spatiotemporal light field describing the illumination $L$ at each time instance $t$ and each spatial location $\bm{x}$ and incident direction $\bm{d}$. This light field is represented by a multilayer perceptron network (MLP) that approximates a six-dimensional (6D) light field function $L(\bm{x}, t,\bm{d})$. The challenge lies in effectively training this MLP from a single image or video. 

Our overarching idea is to leverage 2D diffusion priors for such a 6D MLP training,
inspired by 3D generation using 2D diffusion 
priors~\cite{poole2022dreamfusion,wu2024reconfusion,gao2024cat3d}.
Our 2D diffusion model follows DiffusionLight~\cite{Phongthawee2023DiffusionLight}, which formulates static lighting estimation as an inpainting task: It inserts a chrome ball as a light probe onto the image using a fine-tuned image diffusion model and then unwraps the inpainted chrome ball into an environment map.
However, DiffusionLight is trained to inpaint a single chrome ball at the image center, thus only able to estimate a global environment map at the image's viewpoint. It cannot estimate coherent lighting across multiple spatial locations. 

Instead, we directly train a diffusion model for spatially consistent lighting prediction.
This is done by jointly inpainting multiple chrome balls conditioned on the relative depths of the balls and the background scene.
To this end, we build our training dataset on Infinigen Indoors~\cite{infinigen2024indoors}\textemdash a procedural indoor scene generator based on Blender~\cite{blender}\textemdash by rendering ground-truth environment maps at different spatial locations.
After obtaining the 2D diffusion model, we use it to train an MLP-represented light field $L(\bm{x},t, \bm{d})$. The training process encourages 2D renderings of chrome balls under the light field to follow image priors encoded in the 2D diffusion model.
For a single image input, we simply drop the input time from the MLP and follow the same training procedure to obtain a spatially varying light field.

We evaluate our method on the task of indoor lighting estimation from a single image/video and show superior performance over compared methods, especially when the lighting condition varies spatially and temporally. 
Moreover, we highlight results on spatiotemporal lighting estimation from in-the-wild videos, which remains elusive in previous works.
\section{Related Work}

\paragraph{Lighting Estimation}
Estimating lighting conditions has been a long-standing problem in computer vision and graphics.
Some existing works make use of light probes in the image and perform lighting estimation via inverse rendering~\cite{debevec2008rendering,park2020seeing,yu2023accidental,yi2018faces,verbin2024eclipse}.
Others do not require such light probes to appear in images and often rely on a neural network to predict the lighting.
Many of these works predict single fixed lighting of the scene, either producing an environment map at one particular location where the virtual object is inserted~\cite{liang2025photorealistic} or recovering a full HDR panorama from the input image's limited field of view~\cite{gardner2017learning,yu2019inverserendernet,legendre2019deeplight,somanath2021hdr,wang2022stylelight,dastjerdi2023everlight,Phongthawee2023DiffusionLight}.

Recently, \citet{Phongthawee2023DiffusionLight} proposed DiffusionLight, which uses a pre-trained large-scale image diffusion model to inpaint a chrome ball into the center of an input image. The resulting chrome ball image is then unwrapped into an environment map. This method fine-tunes the diffusion model on a dataset of HDR panorama paired with random crops. Our diffusion model follows a similar design but instead jointly inpaints multiple chrome balls, in order to achieve consistent lighting prediction across different spatial locations of the scene.
To train the model to reason about spatially varying lighting, we construct a synthetic dataset with ground-truth lighting at different spatial locations.

Apart from estimating a single environment map, several recent works can predict spatially varying lighting from a single image by outputting either per-pixel lighting~\cite{garon2019fast,li2020inverse,li2021openrooms,zhu2022irisformer} or a 3D lighting volume~\cite{srinivasan2020lighthouse,wang2021learning,li2023spatiotemporally,wang2024lightoctree}.
In particular, the work by \citet{li2023spatiotemporally} is among the first to take a video input and use a recurrent neural network (RNN) to improve lighting prediction while preserving spatiotemporal consistency. However, this method assumes the input video to capture a static scene and cannot handle dynamic scenes with time-varying lighting conditions.
To our knowledge, our work is the first toward spatiotemporally consistent lighting estimation from a video that captures dynamic lighting.

\paragraph{3D from 2D generative priors}
Recent research have attempted to leverage 2D image diffusion models for 3D content generation from a text prompt~\cite{poole2022dreamfusion,lin2023magic3d,wang2023score} or images~\cite{liu2023zero,shi2023mvdream,gao2024cat3d}. This is motivated by the fact that native 3D data are too limited (in comparison to 2D image data) to train large diffusion models. A pioneer work along this direction, DreamFusion~\cite{poole2022dreamfusion}, proposes score distillation sampling (SDS), wherein a 3D model is optimized with supervision from a 2D image diffusion model. Subsequent works~\cite{zhou2023sparsefusion,wu2024reconfusion,tang2024gaf} all use a variant of SDS, i.e., sampling an image by running multiple DDIM~\cite{song2020denoising} sampling steps on a noisy encoding of the current rendered images. Here, the sampled images serves as a pseudo ground-truth for 3D reconstruction, and the 2D image diffusion model is to provide a prior of partial observations (\emph{i.e.}, the rendered images) of the 3D model.

Our framework follows a similar spirit when optimizing the spatiotemporal light field. Our inpainting diffusion model is meant to provide a prior of partial observations\textemdash in our case, perfectly reflective chrome balls\textemdash of the environment lighting.

\section{Method}

\begin{figure*}
    \centering
    \includegraphics[width=0.99\linewidth]{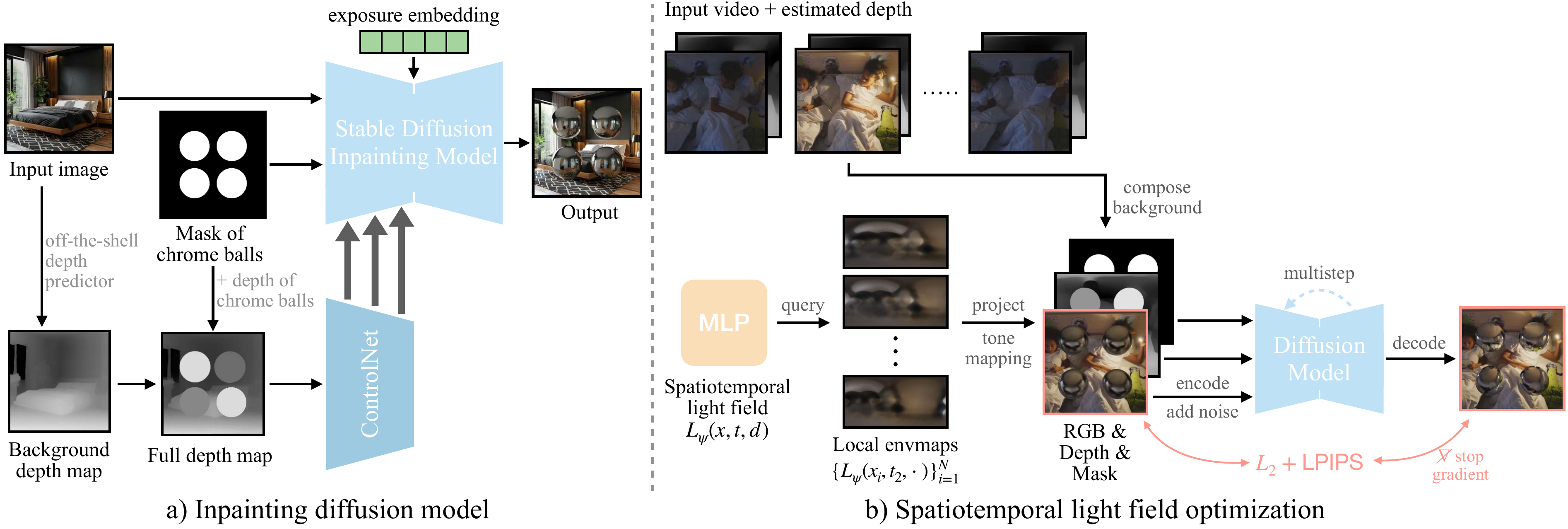} 
    \vspace{-10pt}
    \caption{Method overview. \textbf{(a)} We fine-tune a pre-trained Stable Diffusion Inpainting Model~\cite{Rombach_2022_CVPR} for single image lighting estimation by inpainting multiple chrome balls as light probes in the image (\secref{method_model}). \textbf{(b)} Given a video of an indoor scene with dynamic lighting changes, we distill a spatiotemporal light field (represented as an MLP) from our diffusion model via optimization (\secref{method_opt}). }
    \vspace{-5pt}
    \label{fig:overview}
\end{figure*}

Given an input LDR image (or video) of an indoor scene, we aim to estimate spatially (and temporally) varying HDR lighting represented as environment maps.
Our method starts by fine-tuning a pre-trained image diffusion model to jointly predict lighting at multiple positions (\secref{method_model}).
Then, taking advantage of the learned priors from the diffusion model, we distill a spatially (and temporally) varying light field $L(\bm{x},t,\bm{d})$ represented as an MLP (\secref{method_opt}).
Please refer to \figref{overview} for an overview of our framework.

\subsection{Diffusion Model for Lighting Prediction}\label{sec:method_model}
First, we consider a single LDR image as input. We aim to learn a prior for lighting estimation at locations within the image's view frustum. This prior will be used when presented with a video input for estimating spatiotemporally dynamic lighting (in \secref{method_opt}). 

Inspired by DiffusionLight~\cite{Phongthawee2023DiffusionLight}, we formulate the lighting estimation problem as an inpainting task, \emph{i.e.}, inserting chrome balls onto the image using an image diffusion model and unwrapping them into equirectangular environment maps.
Formally, given an input image $\image$, we estimate the lighting incident to the target locations $\{\bm{x}_i\}_{i=1}^N$. First, we construct a set of chrome balls $B=\{(\bm{x}_i,r_i)\}_{i=1}^N$ as light probes at these locations.
The radius $r_i$ of each chrome ball is chosen such that the diameter of its projection on the image plane is about $1/4$ of the image size.
We use a diffusion model to learn the conditional distribution over the image $\image^B$ with the chrome balls inserted, \emph{i.e.}, $p(\image^B|\image, B)$.

\paragraph{Diffusion Model}
We build our diffusion model on the pre-trained Stable Diffusion inpainting model~\cite{Rombach_2022_CVPR} with depth-conditioned ControlNet~\cite{zhang2023adding}. 
To this end, we first predict the camera intrinsics $K$ and depth map $D_\image$ from the image using off-the-shelf estimators~\cite{jin2023perspective,fspy,ranftl2021vision}.
We then project the chrome balls $B=\{(\bm{x}_i,r_i)\}_{i=1}^N$ onto the image plane to obtain an inpainting mask $\maskBall$, and compose the projected depth of the chrome balls onto the background depth map $D_\image$ to obtain a depth map with chrome balls inserted $\depthBall$. Next, we encode the input image $\image$ via a VAE encoder while resizing the mask $\maskBall$ to the same resolution as the encoded image. 
The encoded $\image$ and $\maskBall$ are then concatenated with the noisy latent vector $z_\tau$ and fed to the denoising U-Net $\denoiseNet$.
At the same time,
the conditional depth map $\depthBall$ is provided as input to the ControlNet. See \figref{overview}-a for an illustration of our model architecture.

To allow for prediction of HDR chrome balls, we follow DiffusionLight~\cite{Phongthawee2023DiffusionLight} to condition the diffusion model on an exposure level \exposure: the illuminance of the inpainted chrome balls in output image $\image^B$ is scaled by $2^{\exposure}$ before being tone-mapped to an LDR image.
Specifically, we condition the diffusion model on an exposure embedding, which is an interpolation of two text CLIP embeddings~\cite{radford2021learning} as a function of \exposure:
\begin{equation}
    \emb^{\exposure} = \emb^{\max} + \frac{\exposure}{\exposure^{\min}}\cdot(\emb^{\min} - \emb^{\max})
\end{equation}
where $\exposure\in[\exposure^{\min}, 0]$.
$\emb^{\max}$ is the embedding of the prompt ``perfect mirrored reflective chrome ball spheres'' while $\emb^{\min}$ is the embedding of the prompt ``perfect black dark mirrored reflective chrome ball spheres''.

We train the model with the standard diffusion training loss on masked pixels:
\begin{equation}
    \vspace{-5pt}
    \mathcal{L} = \mathbb{E}_{z_0,t,\epsilon,\maskBall,\depthBall,\exposure}\left\|\maskBall\odot(\denoiseNet(z_\tau, t, \maskBall,\depthBall,\exposure) - \epsilon)\right\|^2_2
\end{equation}
where $z_\tau$ is the latent vector of inpainted image $\image^B$ at diffusion timestep $\tau$. 
Note that instead of using LoRA as in DiffusionLight \cite{Phongthawee2023DiffusionLight}, we fine-tune all weights of the denoising U-Net and the ControlNet, which we found empirically gives better results.

\paragraph{Dataset Curation}
Training our lighting prediction model requires ground-truth lighting at different spatial locations. Therefore, we construct our training dataset using synthetic indoor scenes.
Among existing datasets of synthetic 3D scenes~\cite{li2021openrooms,infinigen2024indoors,li2022phyir,fu20213d,li2018interiornet}, we choose the recently released Infinigen Indoors \cite{infinigen2024indoors}\textemdash a Blender-based procedural scene generator\textemdash to leverage its diverse lighting conditions of the generated scenes and realistic rendering provided by Blender. 

First, we use Infinigen Indoors to randomly generate 500 indoor scenes, and sample $5$ distinct viewpoints within each scene. 
Each viewpoint is randomly sampled so that the minimum depth from this viewpoint is greater than a certain threshold. This is to ensure that the sampled camera (viewpoint) is not blocked by nearby objects.
\begin{figure*}[!t]
    \centering
    \includegraphics[width=0.99\linewidth]{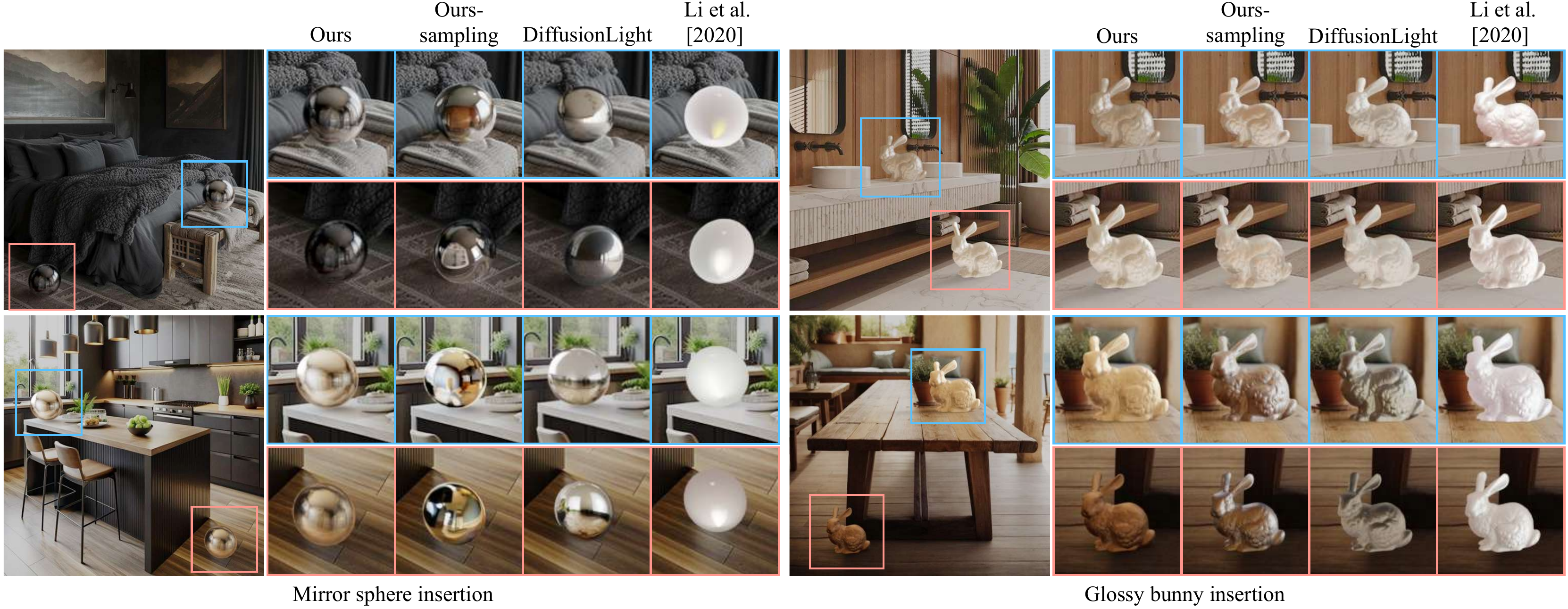}
    \vspace{-10pt}
    \caption{Qualitative comparison of single image lighting prediction for object insertion on in-the-wild scenes. For each example, we insert the object at two different locations. We show the full image of our result on the left and zoom-in crops for each method on the right.}
    \vspace{-10pt}
    \label{fig:comparison_real}
\end{figure*}

Then, for each selected viewpoint, we place $N\sim\mathcal{U}[1,9]$ perfectly reflective chrome balls inside the view frustum
while ensuring that these balls have no intersection with the scene geometry. 
After constructing the scenes, we render HDR images $\image^B$ with chrome balls inserted into the scenes, and also render chrome-ball masks $\maskBall$ and depth maps $\depthBall$.
At the same time, we hide the chrome balls and render the background images $\image$ that serve as paired input.

\subsection{Distilling Spatiotemporal Light Field} \label{sec:method_opt} 
The trained diffusion model can predict HDR lighting at multiple spatial locations from a single image. 
Our ultimate goal is to distill from this 2D diffusion model a spatiotemporally consistent light field 
$\mlp(\bm{x}, t, \bm{d})$,
where $\bm{x}$ is a spatial location, $t\in[1, T]$ is the frame index in time, and $\bm{d}=(\theta, \phi)$ represents an incoming light direction.

Here, we consider a video input $\{\image_t\}_{t=1}^T$ (with $T$ frames) that may capture spatially and temporally varying lighting.
We represent the light field $\mlp(\bm{x}, t, \bm{d})$ using an MLP, and denote
$\mlp(\bm{x}, t, \cdot)$ as the HDR environment map at location $\bm{x}$ and time $t$, one that is obtained by querying the MLP with all incoming light directions.

We view the trained diffusion network as a model that provides partial observations of the scene lighting. From this perspective, those partial observations can be used to supervise the training of the light-field MLP.
A similar view has been taken in training text-to-3D generation: a 3D scene can be distilled from a 2D diffusion model using score distillation sampling (SDS)~\cite{poole2022dreamfusion}.

\begin{table*}[t] 
    \centering
    \caption{Quantitative comparison of single image lighting prediction on synthetic datasets Infinigen Indoor (in-distribution) and 3D-FRONT (out-of-distribution), and and real-world Laval Indoor Spatially Varying (out-of-distribution, real-world).}
    \vspace{-8pt}
    \resizebox{\textwidth}{!}{
    \begin{tabular}{llccccccccc}
    \toprule
    \multirow{2}{*}{Dataset} & \multirow{2}{*}{Method} & 
    \multicolumn{3}{c}{Scale-invariant RMSE $\downarrow$} & 
    \multicolumn{3}{c}{Angular Error $\downarrow$} & 
    \multicolumn{3}{c}{Normalized RMSE $\downarrow$} \\ 
    & & Diffuse & Matte & Mirror & Diffuse & Matte & Mirror & Diffuse & Matte & Mirror \\
    \midrule

    \multirow{5}{*}{\makecell[l]{Infinigen Indoor\\ \cite{infinigen2024indoors}}} 
& DiffusionLight        &                      0.50 &                      0.52 &                      0.58 &                      6.21 &                      6.39 &                      6.62 &                      0.47 &                      0.46 &                      0.46 \\
& DiffusionLight-Distilled        &                      0.60 &                      0.61 &                      0.65 &                      5.85 &                      5.96 &                      \cellcolor{tabsecond}6.11 &                      0.57 &                      0.49 &                      0.48 \\
& Ours-Sampling         &                      0.48 & \cellcolor{tabsecond}0.49 & \cellcolor{tabsecond}0.56 & \cellcolor{tabsecond}5.63 & \cellcolor{tabsecond}5.86 & 6.23 & \cellcolor{tabsecond}0.46 &  \cellcolor{tabfirst}0.42 &  \cellcolor{tabfirst}0.36 \\
& \citet{li2020inverse} & \cellcolor{tabsecond}0.47 &                      0.50 & \cellcolor{tabsecond}0.56 &                      6.50 &                      6.68 &                      6.88 &                      0.62 &                      0.54 &                      0.52 \\
& Ours                  &  \cellcolor{tabfirst}0.41 &  \cellcolor{tabfirst}0.43 &  \cellcolor{tabfirst}0.48 &  \cellcolor{tabfirst}3.72 &  \cellcolor{tabfirst}4.35 &  \cellcolor{tabfirst}4.55 &  \cellcolor{tabfirst}0.45 & \cellcolor{tabsecond}0.44 & \cellcolor{tabsecond}0.45 \\

    \midrule
                                          
    \multirow{5}{*}{\makecell[l]{3D-FRONT\\ \cite{fu20213d}}}
& DiffusionLight        &                      0.32 &                      0.33 &                      0.36 &                      4.95 &                      5.13 &                      5.36 &                      0.47 & \cellcolor{tabsecond}0.47 &                      0.48 \\
& DiffusionLight-Distilled        &                      0.49 &                      0.50 &                      0.52 &                      4.84 &                      5.00 &                      5.17 &                      0.57 &                      0.52 &                      0.52 \\
& Ours-Sampling         & \cellcolor{tabsecond}0.29 & \cellcolor{tabsecond}0.30 & \cellcolor{tabsecond}0.34 & \cellcolor{tabsecond}4.73 & \cellcolor{tabsecond}4.88 & \cellcolor{tabsecond}5.16 & \cellcolor{tabsecond}0.46 &                      0.48 & \cellcolor{tabsecond}0.44 \\
& \citet{li2020inverse} &                      0.30 &                      0.32 &                      0.35 &                      5.73 &                      5.89 &                      6.07 &                      0.62 &                      0.51 &  \cellcolor{tabfirst}0.43 \\
& Ours                  &  \cellcolor{tabfirst}0.25 &  \cellcolor{tabfirst}0.27 &  \cellcolor{tabfirst}0.31 &  \cellcolor{tabfirst}3.49 &  \cellcolor{tabfirst}3.84 &  \cellcolor{tabfirst}4.04 &  \cellcolor{tabfirst}0.43 &  \cellcolor{tabfirst}0.45 &                      0.46 \\
    \midrule
                                          
    \multirow{4}{*}{\makecell[l]{Laval Indoor-Spatially Varying\\ \cite{garon2019fast}}}
& DiffusionLight  &  0.40 &  0.42 &  0.43 &  \cellcolor{tabsecond}7.51 &  7.86 &  8.03 &  \cellcolor{tabsecond}0.53 & 0.60 &  0.65 \\
& Ours-Sampling         & \cellcolor{tabsecond}0.34 & \cellcolor{tabsecond}0.34 & \cellcolor{tabsecond}0.39 & 7.73 & \cellcolor{tabsecond}7.75 & \cellcolor{tabsecond}7.96 & 0.56 &  0.58 & 0.65 \\
& \citet{li2020inverse}  &  0.37 &  \cellcolor{tabsecond}0.34 &  0.42 &  8.27 &  8.42 &  8.76 &  0.63 &  \cellcolor{tabsecond}0.56 &  \cellcolor{tabsecond}0.56 \\
& Ours                  &  \cellcolor{tabfirst}0.26 &  \cellcolor{tabfirst}0.28 &  \cellcolor{tabfirst}0.31 &  \cellcolor{tabfirst}5.45 &  \cellcolor{tabfirst}5.57 &  \cellcolor{tabfirst}6.02 &  \cellcolor{tabfirst}0.48 &  \cellcolor{tabfirst}0.49 &                      \cellcolor{tabfirst}0.53 \\

    \bottomrule
    \end{tabular}%
    }
    \label{tab:comparison}
\end{table*}

\begin{figure*}[!t]
    \centering
    \includegraphics[width=0.99\linewidth]{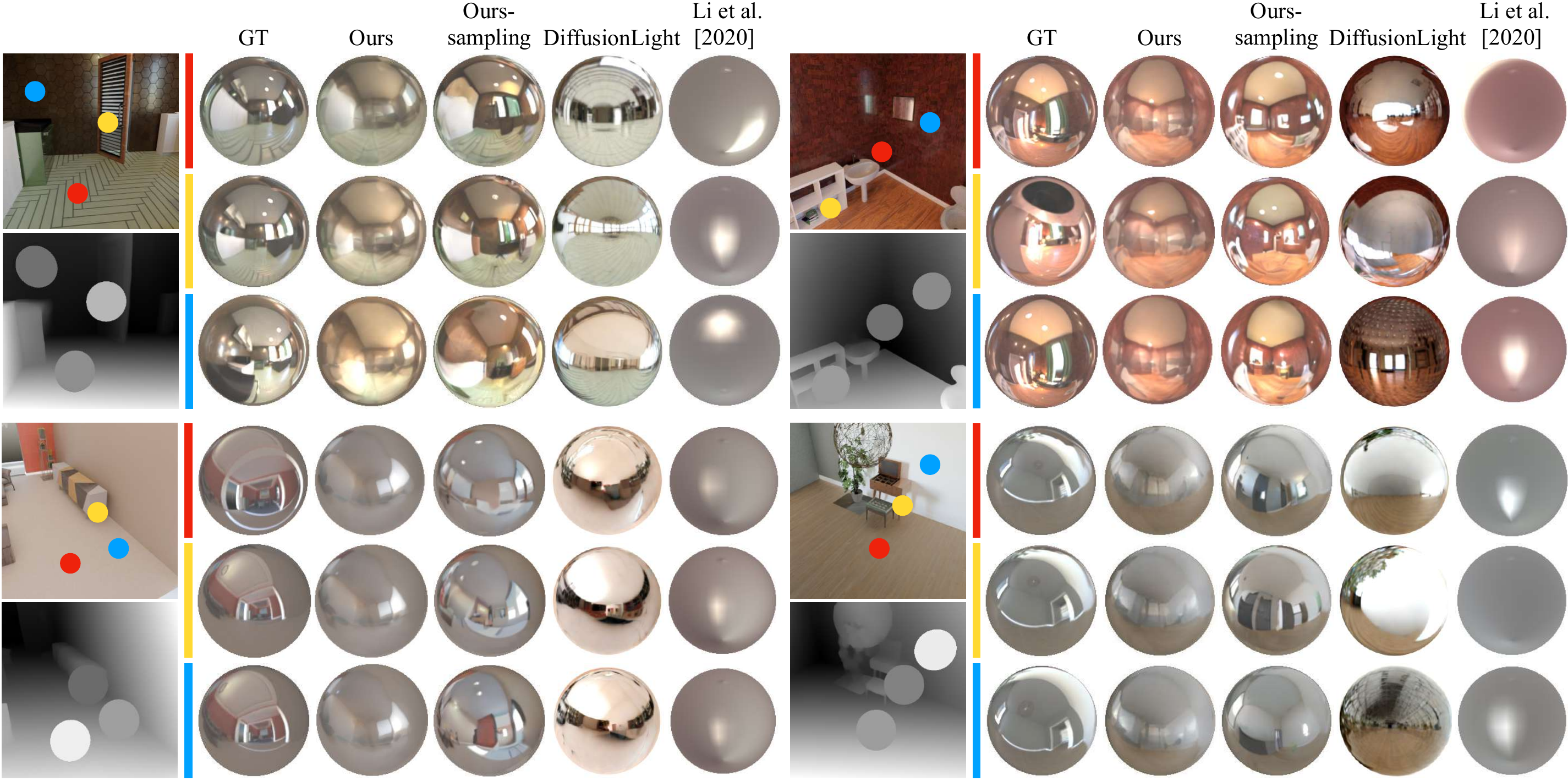}
    \vspace{-8pt}
    \caption{Qualitative comparison of single image indoor lighting prediction on Infinigen Indoor (top) and 3D-FRONT (bottom). For each example, we show on the left the input image and corresponding depth map. Three different locations are marked on the image with color red, yellow and blue. On the right, we show estimated lighting (represented as chrome ball) from each method at the marked locations correspondingly.}
    \vspace{-5pt}
    \label{fig:comparison_eval}
\end{figure*}

In our case, we iteratively improve the MLP. In each iteration,
we randomly pick a frame $\image_t$ and randomly sample $N$ locations within the view frustum to insert chrome balls $B=\{(\bm{x}_i,r_i)\}_{i=1}^N$. With these chrome balls, 
we query the MLP to obtain the HDR environment maps at their locations $\{\mlp(\bm{x}_i, t, \cdot)\}_{i=1}^N$.
We then warp the environment maps onto the chrome balls and project them onto the selected image frame $I_t$.
This process results in a LDR image $\image_t^B$ (which is tone mapped using a sampled exposure value $\exposure$).
The goal of each training iteration is to improve the MLP so that the resulting image $\image_t^B$ better align with the image priors
provided by the 2D diffusion model.


%

In particular, we encode each image in $\image_t^B$ into a latent vector and perturb it into a noisy latent $z_\tau$ with
a noise level $\tau\sim\mathcal{U}[\tau_{\min}, \tau_{\max}]$. 
For each $z_\tau$, we generate a sample using the latent diffusion model by running DDIM sampling~\cite{song2020denoising} for $k$ denoising steps and denote the resulting clean latent sample as $z_0$.
Finally, $z_0$ is decoded into an image $\hat\image_t^B$, one that carries the image prior of our previously trained diffusion model.
Therefore, we treat $\hat\image_t^B$ as a pseudo ground truth for supervision:
\begin{equation}
    \psi = \text{argmin}_\psi\;\mathbb{E}_{t, B,\exposure,\tau}\lft[\|\image_t^B -[\hat\image_t^B]_{\cancel{\nabla}}\|_2^2 + \mathcal{L}_p(\image_t^B, [\hat\image_t^B]_{\cancel{\nabla}})\rgt],
\end{equation}
where $\mathcal{L}_p$ is the perceptual distance LPIPS~\cite{zhang2018unreasonable}, and $[\cdot]_{\cancel{\nabla}}$ is the stop gradient operator, which prevents $\hat\image_t^B$ from being affected by the training process, as it is treated as a pseduo ground truth for providing image priors.
This loss function and optimization strategy resemble the one used in several sparse-view 3D reconstruction methods~\cite{zhou2023sparsefusion,wu2024reconfusion,tang2024gaf}, and we empirically found it to work better than score distillation sampling~\cite{poole2022dreamfusion}.

In practice, we sample the chrome ball locations by sampling $N=9$ pixel locations and then unprojecting each to 3D space with a randomly chosen depth smaller than the background scene depth at that pixel.
During optimization, we linearly decrease $\exposure$ from $0$ to $\exposure^{\min}$ instead of randomly sampling it from $[\exposure^{\min}, 0]$.
We empirically found that this linear sweep of $\exposure$ value helps to synthesize overexposed regions more accurately.

In the case of a single image input, we drop the MLP's input time dimension (\emph{i.e.}, using $T=1$) and follow the above distillation procedure as is.




\subsection{Implementation Details}
\begin{table*}[t!] 
    \centering
    \caption{Ablation study. '\ablationMLP' uses a $3\times 3\times3$ discrete grid with trilinear interpolation between them as the lighting representation for optimization. '\ablationProbes' trains the diffusion model to inpaint only a single sphere probe. '\ablationExposure' uniformly samples the exposure value during optimization instead of gradually decreasing it.}
    \vspace{-10pt}
    \resizebox{\textwidth}{!}{
    \begin{tabular}{llccccccccc}
    \toprule
    \multirow{2}{*}{Dataset} & \multirow{2}{*}{Method} & 
    \multicolumn{3}{c}{Scale-invariant RMSE $\downarrow$} & 
    \multicolumn{3}{c}{Angular Error $\downarrow$} & 
    \multicolumn{3}{c}{Normalized RMSE $\downarrow$} \\ 
    & & Diffuse & Matte & Mirror & Diffuse & Matte & Mirror & Diffuse & Matte & Mirror \\
    \midrule
        
    \multirow{4}{*}{\makecell[l]{Infinigen Indoor\\ \cite{infinigen2024indoors}}}
& \ablationMLP      & \cellcolor{tabsecond}0.47 &                      0.49 & \cellcolor{tabsecond}0.55 &                      8.47 &                      8.63 &                      8.99 &  \cellcolor{tabfirst}0.39 &  \cellcolor{tabfirst}0.40 &  \cellcolor{tabfirst}0.44 \\
& \ablationProbes   &  \cellcolor{tabfirst}0.45 & \cellcolor{tabsecond}0.47 &  \cellcolor{tabfirst}0.53 &                      4.74 &                      4.91 &                      5.12 &                      0.49 &                      0.52 &                      0.55 \\
& \ablationExposure &  \cellcolor{tabfirst}0.45 &  \cellcolor{tabfirst}0.46 &  \cellcolor{tabfirst}0.53 & \cellcolor{tabsecond}4.69 & \cellcolor{tabsecond}4.87 & \cellcolor{tabsecond}5.11 & \cellcolor{tabsecond}0.45 & \cellcolor{tabsecond}0.44 & \cellcolor{tabsecond}0.45 \\
& Ours             &  \cellcolor{tabfirst}0.45 & \cellcolor{tabsecond}0.47 &  \cellcolor{tabfirst}0.53 &  \cellcolor{tabfirst}4.54 &  \cellcolor{tabfirst}4.71 &  \cellcolor{tabfirst}4.95 &                      0.47 &                      0.46 &                      0.47 \\

    \midrule
                                          
    \multirow{4}{*}{\makecell[l]{3D-FRONT\\ \cite{fu20213d}}}
& \ablationMLP      &                      0.36 &                      0.38 &                      0.40 &                      8.04 &                      8.18 &                      8.32 & \cellcolor{tabsecond}0.48 &  \cellcolor{tabfirst}0.45 &  \cellcolor{tabfirst}0.47 \\
& \ablationProbes   & \cellcolor{tabsecond}0.31 & \cellcolor{tabsecond}0.33 &  \cellcolor{tabfirst}0.35 & \cellcolor{tabsecond}3.59 &                      3.67 & \cellcolor{tabsecond}3.83 &                      0.53 &                      0.57 &                      0.61 \\
& \ablationExposure & \cellcolor{tabsecond}0.31 & \cellcolor{tabsecond}0.33 & \cellcolor{tabsecond}0.36 &  \cellcolor{tabfirst}3.49 & \cellcolor{tabsecond}3.62 &  \cellcolor{tabfirst}3.79 &                      0.49 &                      0.52 &                      0.51 \\
& Ours             &  \cellcolor{tabfirst}0.30 &  \cellcolor{tabfirst}0.32 &  \cellcolor{tabfirst}0.35 &  \cellcolor{tabfirst}3.49 &  \cellcolor{tabfirst}3.61 &  \cellcolor{tabfirst}3.79 &  \cellcolor{tabfirst}0.46 & \cellcolor{tabsecond}0.49 & \cellcolor{tabsecond}0.49 \\
%


    \bottomrule
    
    \end{tabular}%
    }
    \vspace{-5pt}
    \label{tab:ablation}
\end{table*}
We fine-tune the diffusion model for 15000 iterations with a batch size of 16 and a learning rate of $10^{-5}$.
To enable classifier-free guidance (CFG), we randomly dropout the exposure embedding with a probability of 0.1.
During fine-tuning, we concatenate the latent background image, the noisy latent image, and the chrome ball mask as input to the diffusion model. This setup allows the model to leverage global context while inpainting the masked regions, maintaining high image quality even when multiple chrome balls occlude a large portion of the background.
Our light field MLP shares a similar architecture as \cite{mildenhall2020nerf} and has 6 hidden layers with a hidden dimension of 256 and positional encoding for the input. We also adopt the skip-connection design as \cite{mildenhall2020nerf} and choose the third layer as the skip layer. 
We use a positional encoding frequency of 6 for the position $\bm{x}$, 4 for the time $t$, and 4 for the direction $\bm{d}$.
During optimization, we fix $\tau_{\max}=1.0$ for all steps, and linearly anneal $\tau_{\min}$ from $1.0$ to $0.0$.
Given $\tau\sim\mathcal{U}[\tau_{\min}, \tau_{\max}]$, we sample the denoised image with $k=\lceil 10\cdot\tau \rceil$ steps and a classifier-free guidance scale of 12.5. Within each optimization step, we firstly adjust the queried HDR environment map from MLP using the sampled exposure level, and then tone-map it into a LDR environment map using a fixed gamma of 2.4, following \cite{wang2022stylelight} and \cite{Phongthawee2023DiffusionLight}.
The diffusion model training takes 14 hours on an NVIDIA A6000, and the distillation procedure typically takes 40 minutes for a video of 100 frames.

\section{Experiments}

In this section, we first evaluate our framework on spatially varying lighting estimation from a single image (\secref{results_spatial}),
and then demonstrate results on spatiotemporally varying lighting estimation from a video (\secref{results_temporal}).
Lastly, we present ablation studies on several design choices (\secref{results_ablation}).
We strongly recommend the readers to view the videos in our supplementary material that best demonstrate the spatiotemporal lighting effects.

\subsection{Single Image Indoor Lighting Estimation}
\label{sec:results_spatial}


\paragraph{Evaluation Datasets}
For quantitative comparison, we test our method on both synthetic and real-world datasets. For the synthetic datasets, we use both in-distribution and out-of-distribution ones, \emph{i.e.}, Infinigen Indoors \cite{infinigen2024indoors} and 3D-FRONT\cite{fu20213d}.
From each dataset, we sample 25 distinct scenes and for each scene render an input image from a randomly sampled viewpoint.
Within the frustum of the input image, we sample 10 different locations and obtain the ground truth lighting by rendering a HDR environment map from a panorama camera placed at each location. For real-world testing, we note that the InteriorNet \cite{li2018interiornet} testing set used in prior work~\cite{srinivasan2020lighthouse,wang2024lightoctree} has become inaccessible. As an alternative, we evaluate our method on the Laval Indoor Spatially Varying HDR Dataset \cite{garon2019fast}, which provides four spatially varying ground-truth lighting conditions per scene, captured using DSLR cameras with exposure bracketing, across 20 different testing scenes.

\paragraph{Evaluation Metrics}
We follow the three-sphere evaluation protocol used in previous work~\cite{gardner2017learning,gardner2019deep,wang2022stylelight,Phongthawee2023DiffusionLight}. Specifically, we use the estimated HDR environment map of size $128 \times 256$ pixels to render three spheres with different materials (gray-diffuse, silver-matte and silver-mirror).
To compare the rendered sphere images to those under the ground truth HDR environment map, we follow the literature and adopt three scale-invariant metrics: scale-invariant Root Mean Square Error (si-RMSE) \cite{5459428}, Angular Error \cite{legendre2019deeplight} and normalized RMSE (mapping the 0.1st and 99.9th percentiles to 0 and 1 \cite{marnerides2018expandnet}).
We report the numbers averaged over all samples locations of all scenes.

\paragraph{Compared Baselines}
Since most of recent methods for predicting spatially varying lighting from a single image do not have open-source code~\cite{wang2024lightoctree,liang2025photorealistic,wang2021learning,li2023spatiotemporally}, we compare against the only one we found available \cite{li2020inverse}. 
\citet{li2020inverse} trained a network to regress an environment map of size $16\times32$ for each pixel, \emph{i.e.}, lighting at each visible surface point of the scene. Yet, in our evaluation dataset, the evaluated location could be arbitrary in the 3D view frustum.
Therefore, we take the predicted lighting at the nearest surface point as their prediction.
In addition, we compare our method with three baselines: 1) direct sampling with DiffusionLight~\cite{Phongthawee2023DiffusionLight} 2) optimization with our distillation with DiffusionLight and 3) direct sampling with our diffusion model (\textit{Ours-sampling}), \emph{i.e.}, independently generating environment maps at each evaluated spatial location.
For the two direct sampling baselines, we follow the HDR merging algorithm in DiffusionLight~\cite{Phongthawee2023DiffusionLight} to obtain HDR environment maps.

\begin{figure}[t]              
  \centering

  \includegraphics[width=0.99\columnwidth]{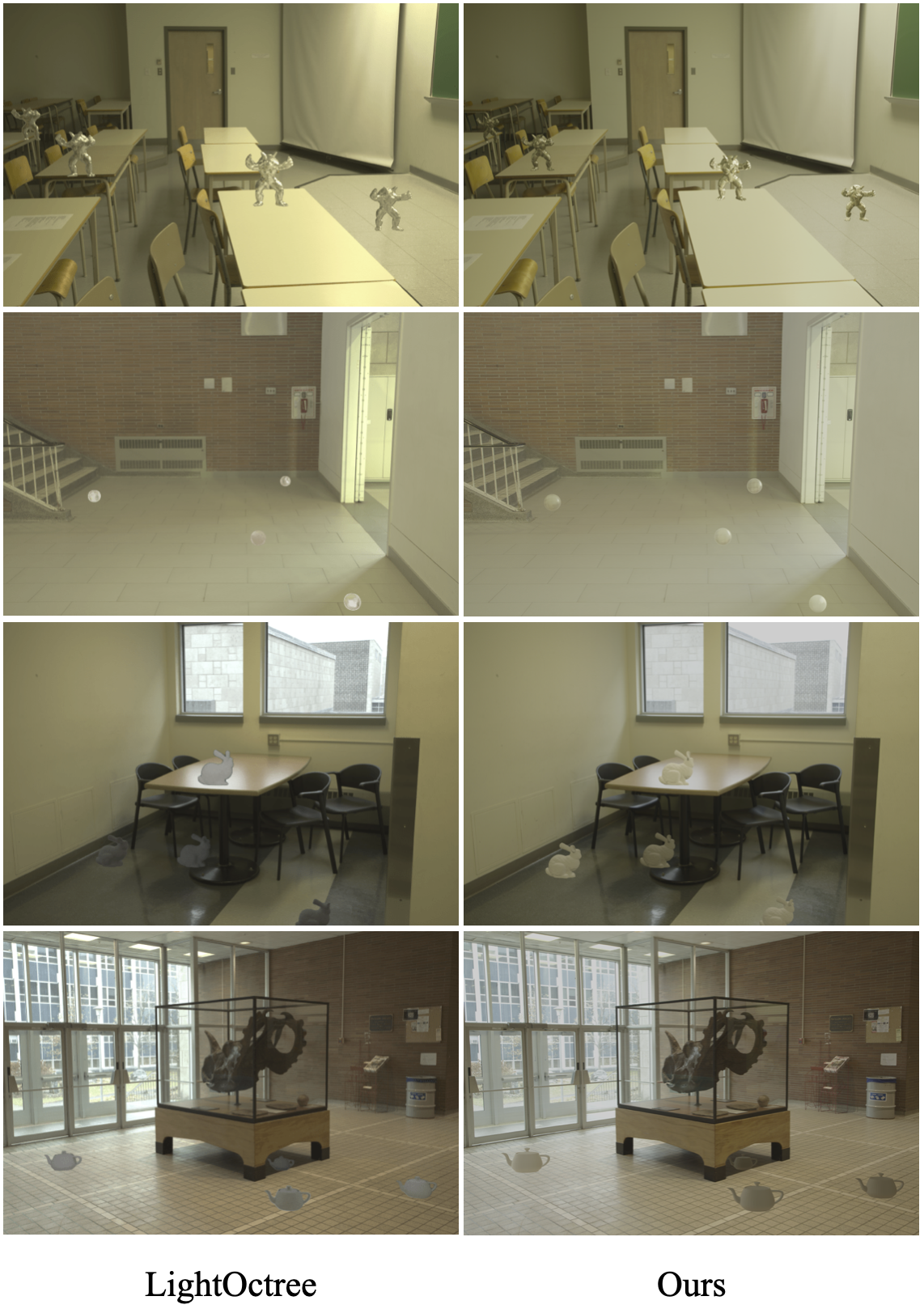}
  \vspace{-10pt}
  \caption{Qualitative evaluation on virtual object insertion on Laval Indoor Spatially Varying HDR dataset~\cite{garon2019fast}. Results of LightOctree~\cite{wang2024lightoctree} are from their paper.}
  \vspace{-10pt}
  \label{fig:comparison_lightoctree}
\end{figure}

\paragraph{Results}
We present quantitative comparison results in \tabref{comparison} and qualitative comparison results in \figref{comparison_eval} and \figref{comparison_lightoctree}.
Our method outperforms compared baselines on almost all evaluation metrics.
While the estimated lighting from \textit{Ours-sampling} and DiffusionLight have sharper texture, they are not consistent across different spatial locations due to independent sampling. The distilled version of DiffusionLight produces spatially smooth lighting estimates but lacks fine-grained detail. This limitation stems from the fact that DiffusionLight samples only a single chrome ball during each optimization step, and its base model is fine-tuned specifically to inpaint chrome balls located at the image center. As a result, it yields increasingly distorted predictions toward the image periphery.

\citet{li2020inverse} does predict spatially varying lighting in a consistent manner, yet their estimated environment map has a limited resolution of $16\times32$ pixels and thus is overly blurred. 
In contrast, our method estimates the spatially varying lighting with sufficient details.
Additionally, we show the comparison results for in-the-wild scenes in \figref{comparison_real} and refer the readers to the videos in the supplementary material that best demonstrate the spatiotemporally varying lighting effects.

In \figref{comparison_lightoctree}, we show qualitative comparison to LightOctree~\cite{wang2024lightoctree} on real world scenes~\cite{garon2019fast}.
Our method achieves comparable results for virtual object insertion.



\subsection{Dynamic Lighting Estimation from a Single Video}
\label{sec:results_temporal}
Our proposed framework enables lighting estimation from a single video of dynamic lighting conditions, which has not been demonstrated in previous work.
As it is not easy to build a synthetic evaluation dataset that reflects realistic lighting changes, we instead choose to present qualitative results on diverse in-the-wild videos in \figref{comparison_video} (and in supplementary material).
We include a baseline where we predict lighting by sampling our diffusion model independently for each frame (\textit{Ours-sampling}).
In comparison, our full pipeline can estimate the scene lighting in a spatiotemporally consistent manner.
We also show object insertion results under dynamic lighting in \figref{teaser} and \figref{teaser_more}.
The visual results are best appreciated in the supplementary videos.


\subsection{Ablation Studies}
\label{sec:results_ablation}

We perform ablation studies by comparing our full pipeline with several variants:
1) \textit{\ablationExposure}, where we uniformly sample the exposure value during optimization instead of gradually decreasing it;
2) \textit{\ablationProbes}, where our diffusion model is trained to inpaint a single sphere probe instead of jointly inpainting multiple probes;
3) \textit{\ablationMLP}, where instead of a MLP we use a $3\times 3\times3$ discrete grid with trilinear interpolation between them as the lighting representation for optimization.
\tabref{ablation} shows the comparison results on a subset of our evaluation dataset for single image indoor lighting estimation.
Our full method surpasses these ablated versions, therefore validating our design choices. We further compare our diffusion model against DiffusionLight to evaluate the performance of the respective base models. Specifically, we test on the Laval Indoor HDR dataset \cite{gardner2017learning}, which differs from the Laval Indoor Spatially Varying dataset in that it provides only a single center lighting ground truth per scene. Following the same one-time central sampling setup used in DiffusionLight, our model achieves si-RMSE values of 0.41/0.43/0.49 for diffuse/matte/mirror spheres, angular errors of 3.61/3.83/4.42, and normalized RMSE values of 0.36/0.34/0.33. These results are comparable to those reported in DiffusionLight (Table 1, line 5, SDXL+LoRA), even though our model is not explicitly trained for this setting, where lighting is captured from the camera viewpoint using a chrome sphere.

\section{Discussion and Limitations}
We present a method that can estimate spatiotemporally consistent lighting from a video of an indoor scene where the illumination varies spatially and temporally.
We highlight our lighting estimation results on challenging real-world videos in the wild, which is rarely demonstrated in the literature.
Our method can also estimate spatially varying lighting from a single image, and we show competitive results compared to prior work.

While our diffusion model generalizes well to in-the-wild indoor scenes thanks to the pre-trained weights, it struggles on outdoor scenes where the sunlight dominates. 
Including synthetic outdoor scenes \cite{raistrick2023infinite} or captured real-world panoramas \cite{hold2019deep} in the training data may help to address this issue.
Our optimization also encounters over-smoothing issues in both the appearance of the reflective spheres and the temporal evolution of lighting, a challenge common to many distillation-based methods \cite{poole2022dreamfusion}. To alleviate this, we experimented with increasing the degree of positional encoding for time t from 4 to 6 and further to 8. While increasing to 6 showed some reduction in over-smoothing, raising the degree to 8 offered only marginal additional improvement and introduced subtle flickering effects temporally. These results suggest diminishing returns beyond a certain encoding complexity. A promising direction for future work is to explore volumetric lighting representations \cite{wang2024lightoctree, srinivasan2020lighthouse}, as volumetric rendering can propagate loss gradients along entire rays, potentially preserving more spatial detail.
Lastly, our method also shares the limitation with many other lighting estimation methods when used for virtual object insertion.
The appearance of rendered object may not seamlessly blend with the image since the graphics renderer is only an approximation of real world light transport.
Adding supervision on the final composed image ~\cite{liang2025photorealistic} could potentially alleviate this problem.

\bibliographystyle{ACM-Reference-Format} 
\bibliography{reference} 


\begin{thebibliography}{50}


\ifx \showCODEN    \undefined \def \showCODEN     #1{\unskip}     \fi
\ifx \showDOI      \undefined \def \showDOI       #1{#1}\fi
\ifx \showISBNx    \undefined \def \showISBNx     #1{\unskip}     \fi
\ifx \showISBNxiii \undefined \def \showISBNxiii  #1{\unskip}     \fi
\ifx \showISSN     \undefined \def \showISSN      #1{\unskip}     \fi
\ifx \showLCCN     \undefined \def \showLCCN      #1{\unskip}     \fi
\ifx \shownote     \undefined \def \shownote      #1{#1}          \fi
\ifx \showarticletitle \undefined \def \showarticletitle #1{#1}   \fi
\ifx \showURL      \undefined \def \showURL       {\relax}        \fi
\providecommand\bibfield[2]{#2}
\providecommand\bibinfo[2]{#2}
\providecommand\natexlab[1]{#1}
\providecommand\showeprint[2][]{arXiv:#2}

\bibitem[Barron and Malik(2014)]%
        {barron2014shape}
\bibfield{author}{\bibinfo{person}{Jonathan~T Barron} {and} \bibinfo{person}{Jitendra Malik}.} \bibinfo{year}{2014}\natexlab{}.
\newblock \showarticletitle{Shape, illumination, and reflectance from shading}.
\newblock \bibinfo{journal}{\emph{IEEE transactions on pattern analysis and machine intelligence}} \bibinfo{volume}{37}, \bibinfo{number}{8} (\bibinfo{year}{2014}), \bibinfo{pages}{1670--1687}.
\newblock


\bibitem[Blender(2024)]%
        {blender}
\bibfield{author}{\bibinfo{person}{Blender}.} \bibinfo{year}{2024}\natexlab{}.
\newblock \bibinfo{booktitle}{\emph{Blender - a 3D modelling and rendering package}}.
\newblock Blender Foundation.
\newblock
\urldef\tempurl%
\url{http://www.blender.org}
\showURL{%
\tempurl}


\bibitem[Dastjerdi et~al\mbox{.}(2023)]%
        {dastjerdi2023everlight}
\bibfield{author}{\bibinfo{person}{Mohammad Reza~Karimi Dastjerdi}, \bibinfo{person}{Jonathan Eisenmann}, \bibinfo{person}{Yannick Hold-Geoffroy}, {and} \bibinfo{person}{Jean-Fran{\c{c}}ois Lalonde}.} \bibinfo{year}{2023}\natexlab{}.
\newblock \showarticletitle{EverLight: Indoor-outdoor editable HDR lighting estimation}. In \bibinfo{booktitle}{\emph{Proceedings of the IEEE/CVF International Conference on Computer Vision}}. \bibinfo{pages}{7420--7429}.
\newblock


\bibitem[Debevec(2008)]%
        {debevec2008rendering}
\bibfield{author}{\bibinfo{person}{Paul Debevec}.} \bibinfo{year}{2008}\natexlab{}.
\newblock \showarticletitle{Rendering synthetic objects into real scenes: Bridging traditional and image-based graphics with global illumination and high dynamic range photography}.
\newblock In \bibinfo{booktitle}{\emph{Acm siggraph 2008 classes}}. \bibinfo{pages}{1--10}.
\newblock


\bibitem[Fu et~al\mbox{.}(2021)]%
        {fu20213d}
\bibfield{author}{\bibinfo{person}{Huan Fu}, \bibinfo{person}{Bowen Cai}, \bibinfo{person}{Lin Gao}, \bibinfo{person}{Ling-Xiao Zhang}, \bibinfo{person}{Jiaming Wang}, \bibinfo{person}{Cao Li}, \bibinfo{person}{Qixun Zeng}, \bibinfo{person}{Chengyue Sun}, \bibinfo{person}{Rongfei Jia}, \bibinfo{person}{Binqiang Zhao}, {et~al\mbox{.}}} \bibinfo{year}{2021}\natexlab{}.
\newblock \showarticletitle{3d-front: 3d furnished rooms with layouts and semantics}. In \bibinfo{booktitle}{\emph{Proceedings of the IEEE/CVF International Conference on Computer Vision}}. \bibinfo{pages}{10933--10942}.
\newblock


\bibitem[Gantelius(2024)]%
        {fspy}
\bibfield{author}{\bibinfo{person}{Per Gantelius}.} \bibinfo{year}{2024}\natexlab{}.
\newblock \bibinfo{booktitle}{\emph{fSpy - an open source, cross platform app for still image camera matching}}.
\newblock
\urldef\tempurl%
\url{https://fspy.io/}
\showURL{%
\tempurl}


\bibitem[Gao et~al\mbox{.}(2024)]%
        {gao2024cat3d}
\bibfield{author}{\bibinfo{person}{Ruiqi Gao}, \bibinfo{person}{Aleksander Holynski}, \bibinfo{person}{Philipp Henzler}, \bibinfo{person}{Arthur Brussee}, \bibinfo{person}{Ricardo Martin-Brualla}, \bibinfo{person}{Pratul Srinivasan}, \bibinfo{person}{Jonathan~T Barron}, {and} \bibinfo{person}{Ben Poole}.} \bibinfo{year}{2024}\natexlab{}.
\newblock \showarticletitle{Cat3d: Create anything in 3d with multi-view diffusion models}.
\newblock \bibinfo{journal}{\emph{arXiv preprint arXiv:2405.10314}} (\bibinfo{year}{2024}).
\newblock


\bibitem[Gardner et~al\mbox{.}(2019)]%
        {gardner2019deep}
\bibfield{author}{\bibinfo{person}{Marc-Andr{\'e} Gardner}, \bibinfo{person}{Yannick Hold-Geoffroy}, \bibinfo{person}{Kalyan Sunkavalli}, \bibinfo{person}{Christian Gagn{\'e}}, {and} \bibinfo{person}{Jean-Fran{\c{c}}ois Lalonde}.} \bibinfo{year}{2019}\natexlab{}.
\newblock \showarticletitle{Deep parametric indoor lighting estimation}. In \bibinfo{booktitle}{\emph{Proceedings of the IEEE/CVF International Conference on Computer Vision}}. \bibinfo{pages}{7175--7183}.
\newblock


\bibitem[Gardner et~al\mbox{.}(2017)]%
        {gardner2017learning}
\bibfield{author}{\bibinfo{person}{Marc-Andr{\'e} Gardner}, \bibinfo{person}{Kalyan Sunkavalli}, \bibinfo{person}{Ersin Yumer}, \bibinfo{person}{Xiaohui Shen}, \bibinfo{person}{Emiliano Gambaretto}, \bibinfo{person}{Christian Gagn{\'e}}, {and} \bibinfo{person}{Jean-Fran{\c{c}}ois Lalonde}.} \bibinfo{year}{2017}\natexlab{}.
\newblock \showarticletitle{Learning to predict indoor illumination from a single image}.
\newblock \bibinfo{journal}{\emph{arXiv preprint arXiv:1704.00090}} (\bibinfo{year}{2017}).
\newblock


\bibitem[Garon et~al\mbox{.}(2019)]%
        {garon2019fast}
\bibfield{author}{\bibinfo{person}{Mathieu Garon}, \bibinfo{person}{Kalyan Sunkavalli}, \bibinfo{person}{Sunil Hadap}, \bibinfo{person}{Nathan Carr}, {and} \bibinfo{person}{Jean-Fran{\c{c}}ois Lalonde}.} \bibinfo{year}{2019}\natexlab{}.
\newblock \showarticletitle{Fast spatially-varying indoor lighting estimation}. In \bibinfo{booktitle}{\emph{Proceedings of the IEEE/CVF Conference on Computer Vision and Pattern Recognition}}. \bibinfo{pages}{6908--6917}.
\newblock


\bibitem[Grosse et~al\mbox{.}(2009)]%
        {5459428}
\bibfield{author}{\bibinfo{person}{Roger Grosse}, \bibinfo{person}{Micah~K. Johnson}, \bibinfo{person}{Edward~H. Adelson}, {and} \bibinfo{person}{William~T. Freeman}.} \bibinfo{year}{2009}\natexlab{}.
\newblock \showarticletitle{Ground truth dataset and baseline evaluations for intrinsic image algorithms}. In \bibinfo{booktitle}{\emph{2009 IEEE 12th International Conference on Computer Vision}}. \bibinfo{pages}{2335--2342}.
\newblock
\urldef\tempurl%
\url{https://doi.org/10.1109/ICCV.2009.5459428}
\showDOI{\tempurl}


\bibitem[Hold-Geoffroy et~al\mbox{.}(2019)]%
        {hold2019deep}
\bibfield{author}{\bibinfo{person}{Yannick Hold-Geoffroy}, \bibinfo{person}{Akshaya Athawale}, {and} \bibinfo{person}{Jean-Fran{\c{c}}ois Lalonde}.} \bibinfo{year}{2019}\natexlab{}.
\newblock \showarticletitle{Deep sky modeling for single image outdoor lighting estimation}. In \bibinfo{booktitle}{\emph{Proceedings of the IEEE/CVF conference on computer vision and pattern recognition}}. \bibinfo{pages}{6927--6935}.
\newblock


\bibitem[Jin et~al\mbox{.}(2023)]%
        {jin2023perspective}
\bibfield{author}{\bibinfo{person}{Linyi Jin}, \bibinfo{person}{Jianming Zhang}, \bibinfo{person}{Yannick Hold-Geoffroy}, \bibinfo{person}{Oliver Wang}, \bibinfo{person}{Kevin Blackburn-Matzen}, \bibinfo{person}{Matthew Sticha}, {and} \bibinfo{person}{David~F Fouhey}.} \bibinfo{year}{2023}\natexlab{}.
\newblock \showarticletitle{Perspective fields for single image camera calibration}. In \bibinfo{booktitle}{\emph{Proceedings of the IEEE/CVF Conference on Computer Vision and Pattern Recognition}}. \bibinfo{pages}{17307--17316}.
\newblock


\bibitem[LeGendre et~al\mbox{.}(2019)]%
        {legendre2019deeplight}
\bibfield{author}{\bibinfo{person}{Chloe LeGendre}, \bibinfo{person}{Wan-Chun Ma}, \bibinfo{person}{Graham Fyffe}, \bibinfo{person}{John Flynn}, \bibinfo{person}{Laurent Charbonnel}, \bibinfo{person}{Jay Busch}, {and} \bibinfo{person}{Paul Debevec}.} \bibinfo{year}{2019}\natexlab{}.
\newblock \showarticletitle{Deeplight: Learning illumination for unconstrained mobile mixed reality}. In \bibinfo{booktitle}{\emph{Proceedings of the IEEE/CVF conference on computer vision and pattern recognition}}. \bibinfo{pages}{5918--5928}.
\newblock


\bibitem[Li et~al\mbox{.}(2018)]%
        {li2018interiornet}
\bibfield{author}{\bibinfo{person}{Wenbin Li}, \bibinfo{person}{Sajad Saeedi}, \bibinfo{person}{John McCormac}, \bibinfo{person}{Ronald Clark}, \bibinfo{person}{Dimos Tzoumanikas}, \bibinfo{person}{Qing Ye}, \bibinfo{person}{Yuzhong Huang}, \bibinfo{person}{Rui Tang}, {and} \bibinfo{person}{Stefan Leutenegger}.} \bibinfo{year}{2018}\natexlab{}.
\newblock \showarticletitle{Interiornet: Mega-scale multi-sensor photo-realistic indoor scenes dataset}.
\newblock \bibinfo{journal}{\emph{arXiv preprint arXiv:1809.00716}} (\bibinfo{year}{2018}).
\newblock


\bibitem[Li et~al\mbox{.}(2020)]%
        {li2020inverse}
\bibfield{author}{\bibinfo{person}{Zhengqin Li}, \bibinfo{person}{Mohammad Shafiei}, \bibinfo{person}{Ravi Ramamoorthi}, \bibinfo{person}{Kalyan Sunkavalli}, {and} \bibinfo{person}{Manmohan Chandraker}.} \bibinfo{year}{2020}\natexlab{}.
\newblock \showarticletitle{Inverse rendering for complex indoor scenes: Shape, spatially-varying lighting and svbrdf from a single image}. In \bibinfo{booktitle}{\emph{Proceedings of the IEEE/CVF Conference on Computer Vision and Pattern Recognition}}. \bibinfo{pages}{2475--2484}.
\newblock


\bibitem[Li et~al\mbox{.}(2022)]%
        {li2022phyir}
\bibfield{author}{\bibinfo{person}{Zhen Li}, \bibinfo{person}{Lingli Wang}, \bibinfo{person}{Xiang Huang}, \bibinfo{person}{Cihui Pan}, {and} \bibinfo{person}{Jiaqi Yang}.} \bibinfo{year}{2022}\natexlab{}.
\newblock \showarticletitle{Phyir: Physics-based inverse rendering for panoramic indoor images}. In \bibinfo{booktitle}{\emph{Proceedings of the IEEE/CVF Conference on Computer Vision and Pattern Recognition}}. \bibinfo{pages}{12713--12723}.
\newblock


\bibitem[Li et~al\mbox{.}(2023)]%
        {li2023spatiotemporally}
\bibfield{author}{\bibinfo{person}{Zhengqin Li}, \bibinfo{person}{Li Yu}, \bibinfo{person}{Mikhail Okunev}, \bibinfo{person}{Manmohan Chandraker}, {and} \bibinfo{person}{Zhao Dong}.} \bibinfo{year}{2023}\natexlab{}.
\newblock \showarticletitle{Spatiotemporally consistent hdr indoor lighting estimation}.
\newblock \bibinfo{journal}{\emph{ACM Transactions on Graphics}} \bibinfo{volume}{42}, \bibinfo{number}{3} (\bibinfo{year}{2023}), \bibinfo{pages}{1--15}.
\newblock


\bibitem[Li et~al\mbox{.}(2021)]%
        {li2021openrooms}
\bibfield{author}{\bibinfo{person}{Zhengqin Li}, \bibinfo{person}{Ting-Wei Yu}, \bibinfo{person}{Shen Sang}, \bibinfo{person}{Sarah Wang}, \bibinfo{person}{Meng Song}, \bibinfo{person}{Yuhan Liu}, \bibinfo{person}{Yu-Ying Yeh}, \bibinfo{person}{Rui Zhu}, \bibinfo{person}{Nitesh Gundavarapu}, \bibinfo{person}{Jia Shi}, {et~al\mbox{.}}} \bibinfo{year}{2021}\natexlab{}.
\newblock \showarticletitle{Openrooms: An open framework for photorealistic indoor scene datasets}. In \bibinfo{booktitle}{\emph{Proceedings of the IEEE/CVF conference on computer vision and pattern recognition}}. \bibinfo{pages}{7190--7199}.
\newblock


\bibitem[Liang et~al\mbox{.}(2025)]%
        {liang2025photorealistic}
\bibfield{author}{\bibinfo{person}{Ruofan Liang}, \bibinfo{person}{Zan Gojcic}, \bibinfo{person}{Merlin Nimier-David}, \bibinfo{person}{David Acuna}, \bibinfo{person}{Nandita Vijaykumar}, \bibinfo{person}{Sanja Fidler}, {and} \bibinfo{person}{Zian Wang}.} \bibinfo{year}{2025}\natexlab{}.
\newblock \showarticletitle{Photorealistic object insertion with diffusion-guided inverse rendering}. In \bibinfo{booktitle}{\emph{European Conference on Computer Vision}}. Springer, \bibinfo{pages}{446--465}.
\newblock


\bibitem[Lin et~al\mbox{.}(2023)]%
        {lin2023magic3d}
\bibfield{author}{\bibinfo{person}{Chen-Hsuan Lin}, \bibinfo{person}{Jun Gao}, \bibinfo{person}{Luming Tang}, \bibinfo{person}{Towaki Takikawa}, \bibinfo{person}{Xiaohui Zeng}, \bibinfo{person}{Xun Huang}, \bibinfo{person}{Karsten Kreis}, \bibinfo{person}{Sanja Fidler}, \bibinfo{person}{Ming-Yu Liu}, {and} \bibinfo{person}{Tsung-Yi Lin}.} \bibinfo{year}{2023}\natexlab{}.
\newblock \showarticletitle{Magic3d: High-resolution text-to-3d content creation}. In \bibinfo{booktitle}{\emph{Proceedings of the IEEE/CVF Conference on Computer Vision and Pattern Recognition}}. \bibinfo{pages}{300--309}.
\newblock


\bibitem[Liu et~al\mbox{.}(2023)]%
        {liu2023zero}
\bibfield{author}{\bibinfo{person}{Ruoshi Liu}, \bibinfo{person}{Rundi Wu}, \bibinfo{person}{Basile Van~Hoorick}, \bibinfo{person}{Pavel Tokmakov}, \bibinfo{person}{Sergey Zakharov}, {and} \bibinfo{person}{Carl Vondrick}.} \bibinfo{year}{2023}\natexlab{}.
\newblock \showarticletitle{Zero-1-to-3: Zero-shot one image to 3d object}. In \bibinfo{booktitle}{\emph{Proceedings of the IEEE/CVF international conference on computer vision}}. \bibinfo{pages}{9298--9309}.
\newblock


\bibitem[Marnerides et~al\mbox{.}(2018)]%
        {marnerides2018expandnet}
\bibfield{author}{\bibinfo{person}{Demetris Marnerides}, \bibinfo{person}{Thomas Bashford-Rogers}, \bibinfo{person}{Jonathan Hatchett}, {and} \bibinfo{person}{Kurt Debattista}.} \bibinfo{year}{2018}\natexlab{}.
\newblock \showarticletitle{Expandnet: A deep convolutional neural network for high dynamic range expansion from low dynamic range content}. In \bibinfo{booktitle}{\emph{Computer Graphics Forum}}, Vol.~\bibinfo{volume}{37}. Wiley Online Library, \bibinfo{pages}{37--49}.
\newblock


\bibitem[Mildenhall et~al\mbox{.}(2020)]%
        {mildenhall2020nerf}
\bibfield{author}{\bibinfo{person}{Ben Mildenhall}, \bibinfo{person}{Pratul~P Srinivasan}, \bibinfo{person}{Matthew Tancik}, \bibinfo{person}{Jonathan~T Barron}, \bibinfo{person}{Ravi Ramamoorthi}, {and} \bibinfo{person}{Ren Ng}.} \bibinfo{year}{2020}\natexlab{}.
\newblock \showarticletitle{NeRF: Representing Scenes as Neural Radiance Fields for View Synthesis}. In \bibinfo{booktitle}{\emph{European Conference on Computer Vision}}. Springer, \bibinfo{pages}{405--421}.
\newblock


\bibitem[Park et~al\mbox{.}(2020)]%
        {park2020seeing}
\bibfield{author}{\bibinfo{person}{Jeong~Joon Park}, \bibinfo{person}{Aleksander Holynski}, {and} \bibinfo{person}{Steven~M Seitz}.} \bibinfo{year}{2020}\natexlab{}.
\newblock \showarticletitle{Seeing the world in a bag of chips}. In \bibinfo{booktitle}{\emph{Proceedings of the IEEE/CVF Conference on Computer Vision and Pattern Recognition}}. \bibinfo{pages}{1417--1427}.
\newblock


\bibitem[Phongthawee et~al\mbox{.}(2024)]%
        {Phongthawee2023DiffusionLight}
\bibfield{author}{\bibinfo{person}{Pakkapon Phongthawee}, \bibinfo{person}{Worameth Chinchuthakun}, \bibinfo{person}{Nontaphat Sinsunthithet}, \bibinfo{person}{Varun Jampani}, \bibinfo{person}{Amit Raj}, \bibinfo{person}{Pramook Khungurn}, {and} \bibinfo{person}{Supasorn Suwajanakorn}.} \bibinfo{year}{2024}\natexlab{}.
\newblock \showarticletitle{Diffusionlight: Light probes for free by painting a chrome ball}. In \bibinfo{booktitle}{\emph{Proceedings of the IEEE/CVF Conference on Computer Vision and Pattern Recognition}}. \bibinfo{pages}{98--108}.
\newblock


\bibitem[Poole et~al\mbox{.}(2022)]%
        {poole2022dreamfusion}
\bibfield{author}{\bibinfo{person}{Ben Poole}, \bibinfo{person}{Ajay Jain}, \bibinfo{person}{Jonathan~T Barron}, {and} \bibinfo{person}{Ben Mildenhall}.} \bibinfo{year}{2022}\natexlab{}.
\newblock \showarticletitle{Dreamfusion: Text-to-3d using 2d diffusion}.
\newblock \bibinfo{journal}{\emph{arXiv preprint arXiv:2209.14988}} (\bibinfo{year}{2022}).
\newblock


\bibitem[Radford et~al\mbox{.}(2021)]%
        {radford2021learning}
\bibfield{author}{\bibinfo{person}{Alec Radford}, \bibinfo{person}{Jong~Wook Kim}, \bibinfo{person}{Chris Hallacy}, \bibinfo{person}{Aditya Ramesh}, \bibinfo{person}{Gabriel Goh}, \bibinfo{person}{Sandhini Agarwal}, \bibinfo{person}{Girish Sastry}, \bibinfo{person}{Amanda Askell}, \bibinfo{person}{Pamela Mishkin}, \bibinfo{person}{Jack Clark}, {et~al\mbox{.}}} \bibinfo{year}{2021}\natexlab{}.
\newblock \showarticletitle{Learning transferable visual models from natural language supervision}. In \bibinfo{booktitle}{\emph{International conference on machine learning}}. PMLR, \bibinfo{pages}{8748--8763}.
\newblock


\bibitem[Raistrick et~al\mbox{.}(2023)]%
        {raistrick2023infinite}
\bibfield{author}{\bibinfo{person}{Alexander Raistrick}, \bibinfo{person}{Lahav Lipson}, \bibinfo{person}{Zeyu Ma}, \bibinfo{person}{Lingjie Mei}, \bibinfo{person}{Mingzhe Wang}, \bibinfo{person}{Yiming Zuo}, \bibinfo{person}{Karhan Kayan}, \bibinfo{person}{Hongyu Wen}, \bibinfo{person}{Beining Han}, \bibinfo{person}{Yihan Wang}, {et~al\mbox{.}}} \bibinfo{year}{2023}\natexlab{}.
\newblock \showarticletitle{Infinite photorealistic worlds using procedural generation}. In \bibinfo{booktitle}{\emph{Proceedings of the IEEE/CVF conference on computer vision and pattern recognition}}. \bibinfo{pages}{12630--12641}.
\newblock


\bibitem[Raistrick et~al\mbox{.}(2024)]%
        {infinigen2024indoors}
\bibfield{author}{\bibinfo{person}{Alexander Raistrick}, \bibinfo{person}{Lingjie Mei}, \bibinfo{person}{Karhan Kayan}, \bibinfo{person}{David Yan}, \bibinfo{person}{Yiming Zuo}, \bibinfo{person}{Beining Han}, \bibinfo{person}{Hongyu Wen}, \bibinfo{person}{Meenal Parakh}, \bibinfo{person}{Stamatis Alexandropoulos}, \bibinfo{person}{Lahav Lipson}, \bibinfo{person}{Zeyu Ma}, {and} \bibinfo{person}{Jia Deng}.} \bibinfo{year}{2024}\natexlab{}.
\newblock \showarticletitle{Infinigen Indoors: Photorealistic Indoor Scenes using Procedural Generation}. In \bibinfo{booktitle}{\emph{Proceedings of the IEEE/CVF Conference on Computer Vision and Pattern Recognition (CVPR)}}. \bibinfo{pages}{21783--21794}.
\newblock


\bibitem[Ranftl et~al\mbox{.}(2021)]%
        {ranftl2021vision}
\bibfield{author}{\bibinfo{person}{Ren{\'e} Ranftl}, \bibinfo{person}{Alexey Bochkovskiy}, {and} \bibinfo{person}{Vladlen Koltun}.} \bibinfo{year}{2021}\natexlab{}.
\newblock \showarticletitle{Vision transformers for dense prediction}. In \bibinfo{booktitle}{\emph{Proceedings of the IEEE/CVF international conference on computer vision}}. \bibinfo{pages}{12179--12188}.
\newblock


\bibitem[Rombach et~al\mbox{.}(2022)]%
        {Rombach_2022_CVPR}
\bibfield{author}{\bibinfo{person}{Robin Rombach}, \bibinfo{person}{Andreas Blattmann}, \bibinfo{person}{Dominik Lorenz}, \bibinfo{person}{Patrick Esser}, {and} \bibinfo{person}{Bj\"orn Ommer}.} \bibinfo{year}{2022}\natexlab{}.
\newblock \showarticletitle{High-Resolution Image Synthesis With Latent Diffusion Models}. In \bibinfo{booktitle}{\emph{Proceedings of the IEEE/CVF Conference on Computer Vision and Pattern Recognition (CVPR)}}. \bibinfo{pages}{10684--10695}.
\newblock


\bibitem[Shi et~al\mbox{.}(2023)]%
        {shi2023mvdream}
\bibfield{author}{\bibinfo{person}{Yichun Shi}, \bibinfo{person}{Peng Wang}, \bibinfo{person}{Jianglong Ye}, \bibinfo{person}{Mai Long}, \bibinfo{person}{Kejie Li}, {and} \bibinfo{person}{Xiao Yang}.} \bibinfo{year}{2023}\natexlab{}.
\newblock \showarticletitle{Mvdream: Multi-view diffusion for 3d generation}.
\newblock \bibinfo{journal}{\emph{arXiv preprint arXiv:2308.16512}} (\bibinfo{year}{2023}).
\newblock


\bibitem[Somanath and Kurz(2021)]%
        {somanath2021hdr}
\bibfield{author}{\bibinfo{person}{Gowri Somanath} {and} \bibinfo{person}{Daniel Kurz}.} \bibinfo{year}{2021}\natexlab{}.
\newblock \showarticletitle{HDR environment map estimation for real-time augmented reality}. In \bibinfo{booktitle}{\emph{Proceedings of the IEEE/CVF Conference on Computer Vision and Pattern Recognition}}. \bibinfo{pages}{11298--11306}.
\newblock


\bibitem[Song et~al\mbox{.}(2020)]%
        {song2020denoising}
\bibfield{author}{\bibinfo{person}{Jiaming Song}, \bibinfo{person}{Chenlin Meng}, {and} \bibinfo{person}{Stefano Ermon}.} \bibinfo{year}{2020}\natexlab{}.
\newblock \showarticletitle{Denoising diffusion implicit models}.
\newblock \bibinfo{journal}{\emph{arXiv preprint arXiv:2010.02502}} (\bibinfo{year}{2020}).
\newblock


\bibitem[Srinivasan et~al\mbox{.}(2020)]%
        {srinivasan2020lighthouse}
\bibfield{author}{\bibinfo{person}{Pratul~P Srinivasan}, \bibinfo{person}{Ben Mildenhall}, \bibinfo{person}{Matthew Tancik}, \bibinfo{person}{Jonathan~T Barron}, \bibinfo{person}{Richard Tucker}, {and} \bibinfo{person}{Noah Snavely}.} \bibinfo{year}{2020}\natexlab{}.
\newblock \showarticletitle{Lighthouse: Predicting lighting volumes for spatially-coherent illumination}. In \bibinfo{booktitle}{\emph{Proceedings of the IEEE/CVF Conference on Computer Vision and Pattern Recognition}}. \bibinfo{pages}{8080--8089}.
\newblock


\bibitem[Tang et~al\mbox{.}(2024)]%
        {tang2024gaf}
\bibfield{author}{\bibinfo{person}{Jiapeng Tang}, \bibinfo{person}{Davide Davoli}, \bibinfo{person}{Tobias Kirschstein}, \bibinfo{person}{Liam Schoneveld}, {and} \bibinfo{person}{Matthias Niessner}.} \bibinfo{year}{2024}\natexlab{}.
\newblock \showarticletitle{GAF: Gaussian Avatar Reconstruction from Monocular Videos via Multi-view Diffusion}.
\newblock \bibinfo{journal}{\emph{arXiv preprint arXiv:2412.10209}} (\bibinfo{year}{2024}).
\newblock


\bibitem[Verbin et~al\mbox{.}(2024)]%
        {verbin2024eclipse}
\bibfield{author}{\bibinfo{person}{Dor Verbin}, \bibinfo{person}{Ben Mildenhall}, \bibinfo{person}{Peter Hedman}, \bibinfo{person}{Jonathan~T Barron}, \bibinfo{person}{Todd Zickler}, {and} \bibinfo{person}{Pratul~P Srinivasan}.} \bibinfo{year}{2024}\natexlab{}.
\newblock \showarticletitle{Eclipse: Disambiguating illumination and materials using unintended shadows}. In \bibinfo{booktitle}{\emph{Proceedings of the IEEE/CVF Conference on Computer Vision and Pattern Recognition}}. \bibinfo{pages}{77--86}.
\newblock


\bibitem[Wang et~al\mbox{.}(2022)]%
        {wang2022stylelight}
\bibfield{author}{\bibinfo{person}{Guangcong Wang}, \bibinfo{person}{Yinuo Yang}, \bibinfo{person}{Chen~Change Loy}, {and} \bibinfo{person}{Ziwei Liu}.} \bibinfo{year}{2022}\natexlab{}.
\newblock \showarticletitle{Stylelight: Hdr panorama generation for lighting estimation and editing}. In \bibinfo{booktitle}{\emph{European Conference on Computer Vision}}. Springer, \bibinfo{pages}{477--492}.
\newblock


\bibitem[Wang et~al\mbox{.}(2023)]%
        {wang2023score}
\bibfield{author}{\bibinfo{person}{Haochen Wang}, \bibinfo{person}{Xiaodan Du}, \bibinfo{person}{Jiahao Li}, \bibinfo{person}{Raymond~A Yeh}, {and} \bibinfo{person}{Greg Shakhnarovich}.} \bibinfo{year}{2023}\natexlab{}.
\newblock \showarticletitle{Score jacobian chaining: Lifting pretrained 2d diffusion models for 3d generation}. In \bibinfo{booktitle}{\emph{Proceedings of the IEEE/CVF Conference on Computer Vision and Pattern Recognition}}. \bibinfo{pages}{12619--12629}.
\newblock


\bibitem[Wang et~al\mbox{.}(2024)]%
        {wang2024lightoctree}
\bibfield{author}{\bibinfo{person}{Xuecan Wang}, \bibinfo{person}{Shibang Xiao}, {and} \bibinfo{person}{Xiaohui Liang}.} \bibinfo{year}{2024}\natexlab{}.
\newblock \showarticletitle{LightOctree: Lightweight 3D Spatially-Coherent Indoor Lighting Estimation}. In \bibinfo{booktitle}{\emph{Proceedings of the IEEE/CVF Conference on Computer Vision and Pattern Recognition}}. \bibinfo{pages}{4536--4545}.
\newblock


\bibitem[Wang et~al\mbox{.}(2021)]%
        {wang2021learning}
\bibfield{author}{\bibinfo{person}{Zian Wang}, \bibinfo{person}{Jonah Philion}, \bibinfo{person}{Sanja Fidler}, {and} \bibinfo{person}{Jan Kautz}.} \bibinfo{year}{2021}\natexlab{}.
\newblock \showarticletitle{Learning indoor inverse rendering with 3d spatially-varying lighting}. In \bibinfo{booktitle}{\emph{Proceedings of the IEEE/CVF International Conference on Computer Vision}}. \bibinfo{pages}{12538--12547}.
\newblock


\bibitem[Wu et~al\mbox{.}(2024)]%
        {wu2024reconfusion}
\bibfield{author}{\bibinfo{person}{Rundi Wu}, \bibinfo{person}{Ben Mildenhall}, \bibinfo{person}{Philipp Henzler}, \bibinfo{person}{Keunhong Park}, \bibinfo{person}{Ruiqi Gao}, \bibinfo{person}{Daniel Watson}, \bibinfo{person}{Pratul~P Srinivasan}, \bibinfo{person}{Dor Verbin}, \bibinfo{person}{Jonathan~T Barron}, \bibinfo{person}{Ben Poole}, {et~al\mbox{.}}} \bibinfo{year}{2024}\natexlab{}.
\newblock \showarticletitle{Reconfusion: 3d reconstruction with diffusion priors}. In \bibinfo{booktitle}{\emph{Proceedings of the IEEE/CVF Conference on Computer Vision and Pattern Recognition}}. \bibinfo{pages}{21551--21561}.
\newblock


\bibitem[Yi et~al\mbox{.}(2018)]%
        {yi2018faces}
\bibfield{author}{\bibinfo{person}{Renjiao Yi}, \bibinfo{person}{Chenyang Zhu}, \bibinfo{person}{Ping Tan}, {and} \bibinfo{person}{Stephen Lin}.} \bibinfo{year}{2018}\natexlab{}.
\newblock \showarticletitle{Faces as lighting probes via unsupervised deep highlight extraction}. In \bibinfo{booktitle}{\emph{Proceedings of the European Conference on computer vision (ECCV)}}. \bibinfo{pages}{317--333}.
\newblock


\bibitem[Yu et~al\mbox{.}(2023)]%
        {yu2023accidental}
\bibfield{author}{\bibinfo{person}{Hong-Xing Yu}, \bibinfo{person}{Samir Agarwala}, \bibinfo{person}{Charles Herrmann}, \bibinfo{person}{Richard Szeliski}, \bibinfo{person}{Noah Snavely}, \bibinfo{person}{Jiajun Wu}, {and} \bibinfo{person}{Deqing Sun}.} \bibinfo{year}{2023}\natexlab{}.
\newblock \showarticletitle{Accidental light probes}. In \bibinfo{booktitle}{\emph{Proceedings of the IEEE/CVF Conference on Computer Vision and Pattern Recognition}}. \bibinfo{pages}{12521--12530}.
\newblock


\bibitem[Yu and Smith(2019)]%
        {yu2019inverserendernet}
\bibfield{author}{\bibinfo{person}{Ye Yu} {and} \bibinfo{person}{William~AP Smith}.} \bibinfo{year}{2019}\natexlab{}.
\newblock \showarticletitle{Inverserendernet: Learning single image inverse rendering}. In \bibinfo{booktitle}{\emph{Proceedings of the IEEE/CVF Conference on Computer Vision and Pattern Recognition}}. \bibinfo{pages}{3155--3164}.
\newblock


\bibitem[Zhang et~al\mbox{.}(2023)]%
        {zhang2023adding}
\bibfield{author}{\bibinfo{person}{Lvmin Zhang}, \bibinfo{person}{Anyi Rao}, {and} \bibinfo{person}{Maneesh Agrawala}.} \bibinfo{year}{2023}\natexlab{}.
\newblock \showarticletitle{Adding conditional control to text-to-image diffusion models}. In \bibinfo{booktitle}{\emph{Proceedings of the IEEE/CVF International Conference on Computer Vision}}. \bibinfo{pages}{3836--3847}.
\newblock


\bibitem[Zhang et~al\mbox{.}(2018)]%
        {zhang2018unreasonable}
\bibfield{author}{\bibinfo{person}{Richard Zhang}, \bibinfo{person}{Phillip Isola}, \bibinfo{person}{Alexei~A Efros}, \bibinfo{person}{Eli Shechtman}, {and} \bibinfo{person}{Oliver Wang}.} \bibinfo{year}{2018}\natexlab{}.
\newblock \showarticletitle{The unreasonable effectiveness of deep features as a perceptual metric}. In \bibinfo{booktitle}{\emph{Proceedings of the IEEE conference on computer vision and pattern recognition}}. \bibinfo{pages}{586--595}.
\newblock


\bibitem[Zhou and Tulsiani(2023)]%
        {zhou2023sparsefusion}
\bibfield{author}{\bibinfo{person}{Zhizhuo Zhou} {and} \bibinfo{person}{Shubham Tulsiani}.} \bibinfo{year}{2023}\natexlab{}.
\newblock \showarticletitle{Sparsefusion: Distilling view-conditioned diffusion for 3d reconstruction}. In \bibinfo{booktitle}{\emph{Proceedings of the IEEE/CVF Conference on Computer Vision and Pattern Recognition}}. \bibinfo{pages}{12588--12597}.
\newblock


\bibitem[Zhu et~al\mbox{.}(2022)]%
        {zhu2022irisformer}
\bibfield{author}{\bibinfo{person}{Rui Zhu}, \bibinfo{person}{Zhengqin Li}, \bibinfo{person}{Janarbek Matai}, \bibinfo{person}{Fatih Porikli}, {and} \bibinfo{person}{Manmohan Chandraker}.} \bibinfo{year}{2022}\natexlab{}.
\newblock \showarticletitle{Irisformer: Dense vision transformers for single-image inverse rendering in indoor scenes}. In \bibinfo{booktitle}{\emph{Proceedings of the IEEE/CVF Conference on Computer Vision and Pattern Recognition}}. \bibinfo{pages}{2822--2831}.
\newblock


\end{thebibliography}

\begin{figure*}
    \centering
    \includegraphics[width=0.94\linewidth]{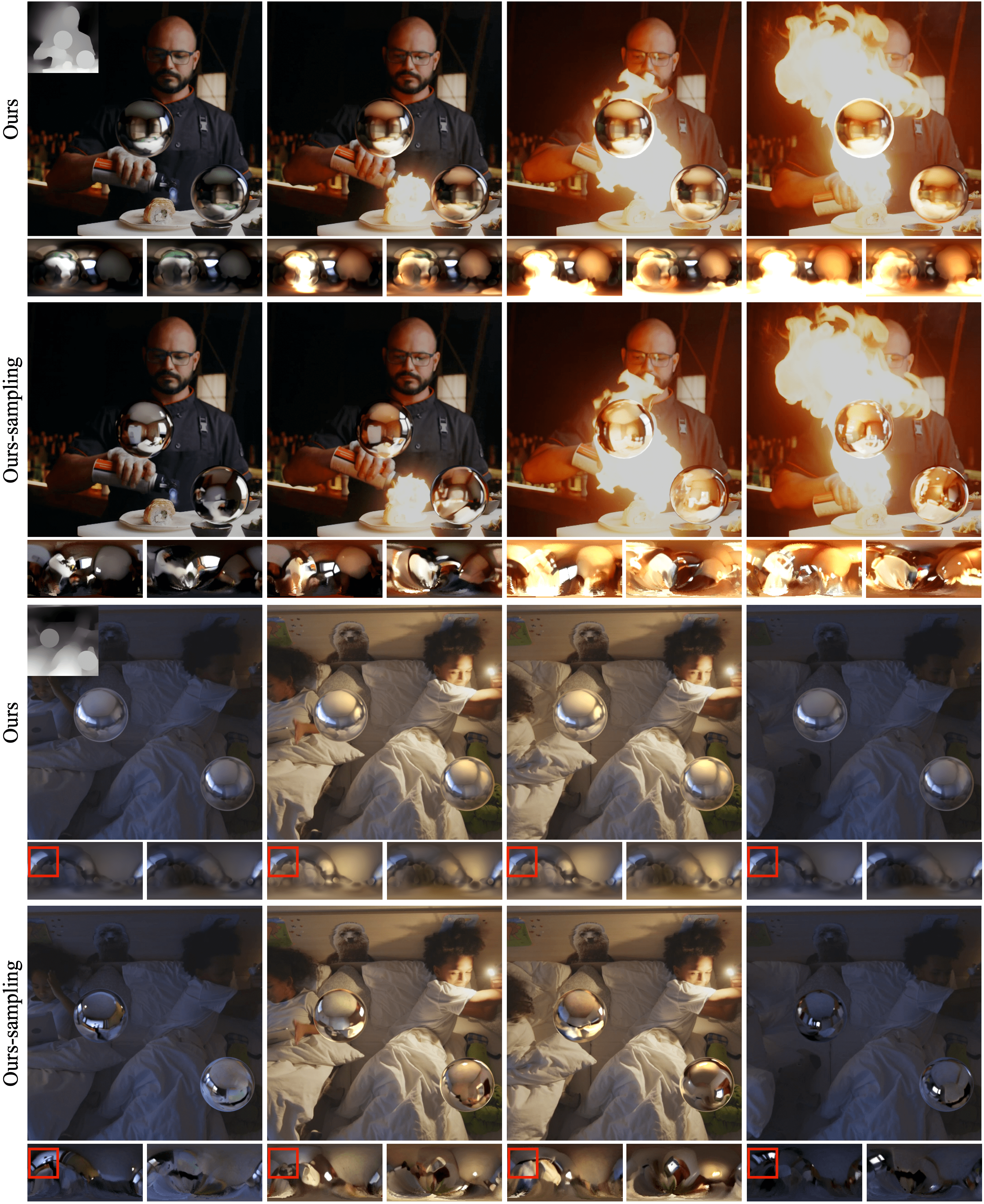}
    \vspace{-10pt}
    \caption{Qualitative results of dynamic lighting estimation from single video on \emph{in-the-wild} scenes. For each example video, we show 4 frames with results from our full pipeline (\textit{Ours}) and our diffusion model samples (\textit{Ours-sampling}). For each frame, we show estimated lighting at two different locations, depicted as chrome balls on the image with corresponding environment maps below. Depth map of the scene is also shown on top left of the first frame. The estimated lighting from our full pipeline has better temporal consistency (see the red box crops). Please refer to the supplement for video results of more examples.}
    \label{fig:comparison_video}
\end{figure*}

\begin{figure*}
    \centering
    \includegraphics[width=0.99\linewidth]{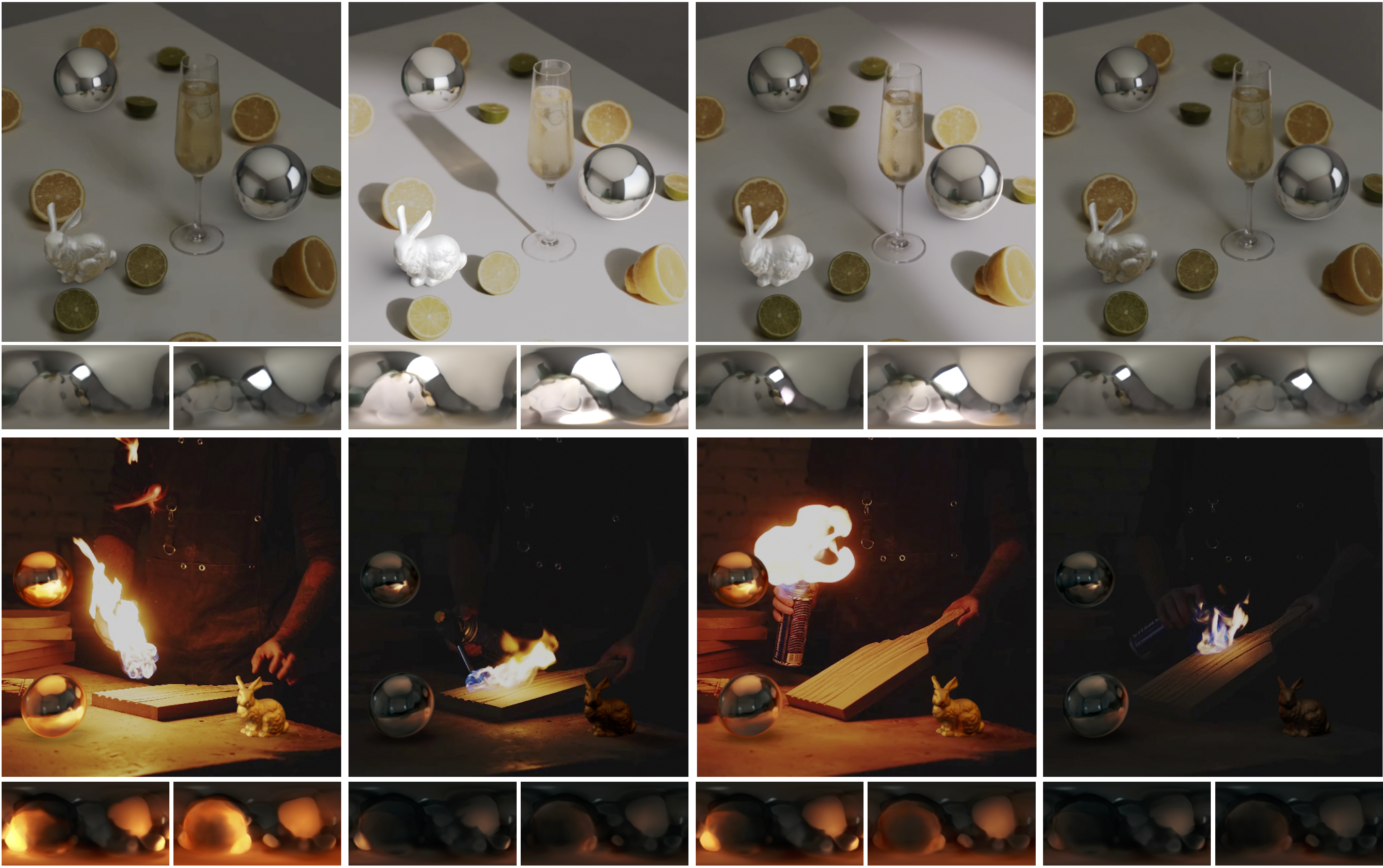}
    \vspace{-2pt}
    \caption{More examples of object insertion under dynamic lighting. Here we show four frames of each input video. Here we show four frames of an video. For each frame, we show \textbf{(top)} the result of virtual object insertion under the estimated lighting and \textbf{(bottom)} estimated environment maps at locations where the two mirror spheres are inserted. In each column, the environment map on the bottom-left corresponds to the top mirror sphere, and the one on the bottom-right corresponds to the bottom mirror sphere.
    }
    \label{fig:teaser_more}
\end{figure*}

\end{document}